\documentclass[preprintnumbers,aps,prd,
notitlepage,
superscriptaddress,nofootinbib]{revtex4-1}
  \usepackage{latexsym,bm,amsmath,amssymb,amsfonts}
\newcommand{\beq}{\begin{eqnarray}}
\newcommand{\eeq}{\end{eqnarray}}

\newcommand{\ttau}{\tilde{\tau}}
\newcommand{\teta}{\tilde{\eta}}

\usepackage{graphicx}
\usepackage{amsmath,amsthm,amssymb}

\begin{document}

\preprint{YITP-15-114}

\title{A non-boost-invariant solution of relativistic hydrodynamics \\ in 1+3 dimensions}

\author{Yoshitaka Hatta}
\affiliation{Yukawa Institute for Theoretical Physics, Kyoto University, Kyoto 606-8502, Japan}

\author{Bo-Wen Xiao}
\affiliation{Key Laboratory of Quark and Lepton Physics (MOE) and Institute
of Particle Physics, Central China Normal University, Wuhan 430079, China}

\author{Di-Lun Yang}
\affiliation{Theoretical Research Division, Nishina Center, RIKEN, Wako, Saitama 351-0198, Japan}

\date{\today}
\vspace{10mm}

\begin{abstract}
 We present a new solution of relativistic hydrodynamics in 1+3 dimensions which depends on both the transverse coordinate and rapidity. At early times the flow expands dominantly longitudinally in a non-boost-invariant manner, and at late times it expands nearly spherically. These two regimes are shown to be related by symmetry. The effect of viscosity is also discussed.
\end{abstract}

\maketitle

\section{Introduction}

There are two well-known solutions of the relativistic hydrodynamic equation which are intended to describe the evolution of the dense matter created in heavy-ion collisions.  In Landau's picture \cite{landau}, the colliding nuclei come to a complete halt and deposit energy in a region extended in the longitudinal (beam) direction. Subsequently, this region expands one-dimensionally into the vacuum due to the large longitudinal gradient. The Khalatnikov-Landau solution \cite{kha,landau} of relativistic hydrodynamics in 1+1 dimensions offers a concrete realization of this idea.
On the other hand, in Bjorken's picture \cite{Bjorken:1982qr}, the highly Lorentz-contracted nuclei pass through each other leaving behind a dense partonic region which expands longitudinally in a boost-invariant (rapidity-independent) manner. This is described by the Bjorken solution \cite{Bjorken:1982qr} and is considered to be a plausible picture at very high energy.

In realistic collisions, boost invariance is violated, and what actually happens is somewhere in between perfect stopping and perfect transparency. To accommodate this, there have been a number of attempts to interpolate the two solutions \cite{Srivastava:1992gh,Pratt:2008jj,Bialas:2007iu,Nagy:2007xn,Beuf:2008vd,Wong:2008ex,Beuf:2008vd,Gubser:2012gy,Wong:2014sda}. However, all of these works essentially deal with 1+1-dimensional hydrodynamics implicitly assuming, rather unrealistically, that the colliding nuclei have infinite transverse extent. As already observed by Landau \cite{landau}, the one-dimensional expansion is eventually superseded by a three-dimensional one when the time reaches of the order of the nuclear transverse size. Yet, obtaining full-fledged 1+3-dimensional analytical solutions is quite challenging because the hydrodynamic equations intimately couple the longitudinal and transverse dynamics.
%\footnote{There exist spherically symmetric exact solutions in 1+3-dimensions. We deem these solutions  essentially 1+1-dimensional, since they depend only on time and the radial coordinate  $r=|\vec{r}|$. In any case, they are not realistic models of heavy-ion collisions.}

A notable exception is the solution obtained by Gubser \cite{Gubser:2010ze}. This generalizes the Bjorken solution by adding transverse expansion while retaining boost invariance. Yet, like the 1+1-dimensional solutions, Gubser's solution depends only on two variables in a cleverly chosen coordinate system. Introducing the dependence on a third variable (rapidity or the azimuthal angle) is difficult and so far has been done in the form of small perturbations \cite{Gubser:2010ui,Hatta:2014jva} (see, however, \cite{Csorgo:2003ry,Csanad:2014dpa,Martinez:2010sd}).

In this paper, we analytically construct a  non-boost-invariant 1+3-dimensional solution of relativistic hydrodynamics which essentially depends on three variables and has appealing features as a model of low energy heavy-ion collisions. Namely, the flow expands dominantly longitudinally in a non-boost-invariant manner at early times $t\ll L$ where $L$ is the size of the nucleus, and at late times $t\gg L$ it expands nearly spherically. This will be demonstrated fully analytically.  In the intermediate regime $t\sim L$, we have not found a closed analytic expression. We however develop a perturbation theory to approach this regime expanding around the early/late-time solutions. At the end, we discuss the effect of viscosity by approximately solving the Navier-Stokes equation.

\section{Hydrodynamics in $dS_2\times H_2$}

Our starting point is the relativistic hydrodynamic equation for an ideal fluid
\beq
u^\mu \partial_\mu \varepsilon + (\varepsilon +p)\nabla_\mu u^\mu=0\,,\qquad
(\varepsilon + p)u^\nu\nabla_\nu u^\mu + \Delta^{\mu\nu}\partial_\nu p=0\,, \label{eq}
\eeq
where $\varepsilon$ is the energy density, $p$ is the pressure and $u^\mu$ is the flow velocity normalized as $u^\mu u_\mu=-1$. $\nabla_\mu$ is the covariant derivative and $\Delta^{\mu\nu}=g^{\mu\nu}+u^\mu u^\nu$ is the projection operator transverse to the flow. We assume the relativistic (conformal) equation of state $p=\frac{1}{3}\varepsilon \propto T^4$ with $T$ being the temperature. For our purpose, it is more convenient to rewrite (\ref{eq}) in a different, but equivalent form
  \beq
 \nabla_\mu (\sigma u^\mu)=0\,, \label{hydro1} \\
  u^\mu\bigl(\nabla_\mu (T u_\nu) -\nabla_\nu (T u_\mu)\bigr) =0\,, \label{hydro2}
 \eeq
 where $\sigma\propto T^3$ is the entropy density.
 In the presence of conserved charges, one should couple the above equations with the continuity equation
 \beq
 \nabla_\mu (n u^\mu)=0\,,\label{con}
 \eeq
 where $n$ is the charge density. However, in ideal hydrodynamics (\ref{con}) is not an independent equation and trivially solved by $n \propto T^3$.

\begin{figure}[tbp]
 \includegraphics[width=120mm,height=90mm]{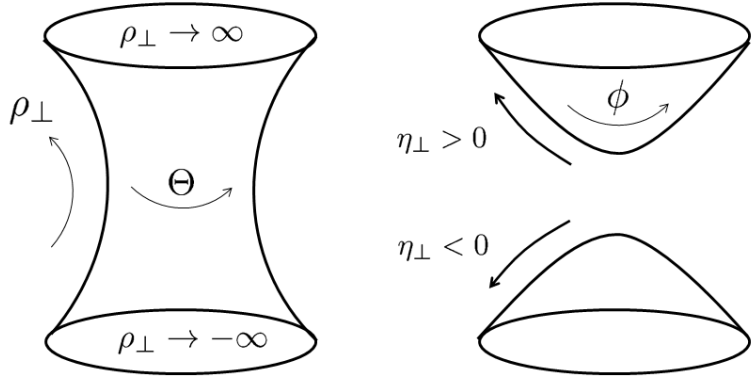}
 \caption{The de Sitter space $dS_2$ (left) and the hyperbolic space $H_2$ (right). $\rho_\perp$ plays the role of `time' in the product space $dS_2\times H_2$. \label{fig1}}
\end{figure}

As demonstrated in \cite{Gubser:2010ze,Hatta:2014gqa,Hatta:2014gga}, if the equation of state is relativistic $\varepsilon=3p$, a powerful method to construct nontrivial solutions is available. Instead of working in Minkowski space with the metric
\beq
ds^2&=&-dt^2+dx^2+dy^2 + dz^2\,,
% \nonumber \\
% &=& -d\tau^2 + dx_\perp^2 + x_\perp^2 d\phi^2 +\tau^2d\eta^2\,, \label{weyl}
 \eeq
%$\tau=\sqrt{t^2-z^2}$ is the proper time and $\eta = \frac{1}{2}\ln \frac{t+z}{t-z} $ is the spacetime rapidity often used in heavy-ion collisions.
 one can work in different coordinate systems which are related to Minkowski space via the Weyl transformation. Consider, then, the following coordinate transformation
\beq
ds^2=-dt^2+dx_\perp^2+x_\perp^2 d\phi^2 + dz^2 &=& -d\tau_\perp^2 + \tau^2_\perp d\eta_\perp^2 + \tau_\perp^2\sinh^2\eta_\perp d\phi^2 + dz^2 \nonumber  \\
&=& \tau_\perp^2 \left( \frac{-d\tau_\perp^2+dz^2}{\tau_\perp^2} + d\eta_\perp^2 + \sinh^2\eta_\perp d\phi^2\right) \,, \label{xp}
\eeq
 where $x_\perp=\sqrt{x^2+y^2}$ and $\phi$ is the azimuthal angle. The `transverse proper time' $\tau_\perp$ and the `transverse rapidity' $\eta_\perp$
 are defined as
 \beq
 \tau_\perp =\sqrt{t^2-x_\perp^2}\,, \qquad \eta_\perp = \frac{1}{2}\ln \frac{t+x_\perp}{t-x_\perp}\,. \label{tran}
 \eeq
 The coordinates ($\tau_\perp,\eta_\perp$) were previously introduced in \cite{Lin:2009kv}. This coordinate system covers only the region $t>x_\perp$. The case $x_\perp >t$ will be treated later.
 In the second line of (\ref{xp}), we observe that the Weyl rescaled metric $ds^2/\tau_\perp^2$ is that of $dS_2\times H_2$, the product of the two-dimensional de Sitter space and the two-dimensional hyperbolic space. Further transforming to the so-called global coordinates of $dS_2$ (see Fig.~\ref{fig1}), we arrive at
 \beq
 d\hat{s}^2\equiv \frac{ds^2}{\tau_\perp^2} =
  -d\rho_\perp^2+\cosh^2\rho_\perp d\Theta^2 + d\eta_\perp^2+\sinh^2\eta_\perp d\phi^2\,, \label{ds}
  \eeq
  where
\beq
\sinh\rho_\perp = \frac{\tau^2_\perp-L^2-z^2}{2L\tau_\perp}\,, \qquad \tan \Theta=\frac{2Lz}{L^2+\tau^2_\perp-z^2}\,. \label{def}
\eeq
 $L$ is an arbitrary length parameter, and can be considered as the transverse size of the nucleus.  We shall work in the coordinates (\ref{ds}) and solve the hydrodynamic equations in the form (\ref{hydro1}) and (\ref{hydro2}). Solutions $\{\hat{u}^\mu,\hat{\varepsilon}\}$ are then transformed back to Minkowski space via the formulas
\beq
u_\mu = \tau_\perp \frac{\partial \hat{x}^\nu}{\partial x^\mu}\hat{u}_\nu\,, \qquad
\varepsilon=\frac{\hat{\varepsilon}}{\tau_\perp^4}\,.
\eeq

As a slight generalization, the above transformation can be combined with the time translation. Instead of (\ref{tran}), we may define
 \beq
 \tau_\perp =\sqrt{(t-t_0)^2-x_\perp^2}\,, \qquad \eta_\perp = \frac{1}{2}\ln \frac{t-t_0+x_\perp}{t-t_0-x_\perp}\,, \label{note}
 \eeq
 where $t_0$ is arbitrary.
 In the following we only show the results for $t_0=0$, but one can freely replace $t$ with $t-t_0$.

\subsection{An exact solution}

Consider the comoving flow in $dS_2\times H_2$
\beq
(\hat{u}^{\rho_\perp},\hat{u}^\Theta,\hat{u}^{\eta_\perp},\hat{u}^\phi)=(1,0,0,0)\,.
\eeq
 With this velocity, (\ref{hydro2}) becomes trivial and (\ref{hydro1}) can be easily solved as
\beq
\hat{\varepsilon}\propto \left(\frac{1}{\cosh\rho_\perp}\right)^{4/3}\,.
\eeq
The solution in Minkowski space is
\beq
(u_t,\vec{u}_\perp,u^z)&=&\frac{1}{\sqrt{
(t^2-x_\perp^2-L^2-z^2)^2+4L^2(t^2-x_\perp^2)}} \nonumber \\
&& \qquad \times \left(
\frac{t(t^2-x_\perp^2+L^2+z^2)}{\sqrt{t^2-x_\perp^2}},\frac{\vec{x}_\perp (t^2-x_\perp^2+L^2+z^2)}{\sqrt{t^2-x_\perp^2}},2z\sqrt{t^2-x_\perp^2}\right)\,,
\eeq
\beq
\varepsilon \propto \frac{1}{(t^2-x_\perp^2)^{4/3}} \left(\frac{1}{4L^2(t^2-x_\perp^2)+(t^2-x_\perp^2-L^2-z^2)^2}\right)^{2/3}\,.
%\nonumber \\
%&=& \frac{1}{(\tau^2\cosh^2\eta-x_\perp^2)^{4/3}} \left(\frac{1}{4L^2(\tau^2\cosh^2\eta-x_\perp^2)+(L^2-\tau^2+x_\perp^2)^2}\right)^{2/3}\,.
\label{ex}
\eeq
This is a new exact solution. It is analogous to Gubser flow by construction  \cite{Gubser:2010ze}, but instead of expanding longitudinally in the $z$-direction, the fluid is expanding radially in the $x_\perp$ direction at the speed of light. It is thus not an attractive model of heavy-ion collisions.

\section{The new solution}

\subsection{Asymptotic solutions}

 We now consider a more general class of solutions of the form
 \beq
 (\hat{u}^{\rho_\perp},\hat{u}^\Theta,\hat{u}^{\eta_\perp},\hat{u}^\phi)
 =(\cosh\alpha,0,\sinh\alpha,0)\,, \label{two}
 \eeq
 where $\alpha=\alpha(\rho_\perp, \eta_\perp)$ is the fluid rapidity in this space.
Since the flow velocity (\ref{two}) has only two components, the solution of (\ref{hydro2}) takes the form of  potential flow in 1+1-dimensions
  \beq
  \hat{T}\hat{u}_{\rho_\perp} = -\hat{T}\cosh\alpha=\partial_{\rho_\perp} \Phi(\rho_\perp,\eta_\perp)\,, \qquad
  \hat{T}\hat{u}_{\eta_\perp}=\hat{T}\sinh\alpha =\partial_{\eta_\perp} \Phi(\rho_\perp,\eta_\perp)\,.
  \label{pot}
  \eeq
In 1+1-dimensions, a standard way to proceed is to introduce the Khalatnikov potential $\chi(\hat{T},\alpha)$ \cite{kha,landau} as the Legendre transform of the potential $\Phi(\rho_\perp,\eta_\perp)$ and solve the resulting equation for $\chi$.
 However, the present problem  does not fully reduce to a 1+1-dimensional one because the metric depends on $\rho_\perp$ and $\eta_\perp$, and this makes the analysis in terms of $\chi$ difficult. Instead, we develop a systematic perturbation theory to determine the function $\alpha(\rho_\perp,\eta_\perp)$ order by order.

\begin{figure}[tbp]
  \includegraphics[width=70mm,height=70mm]{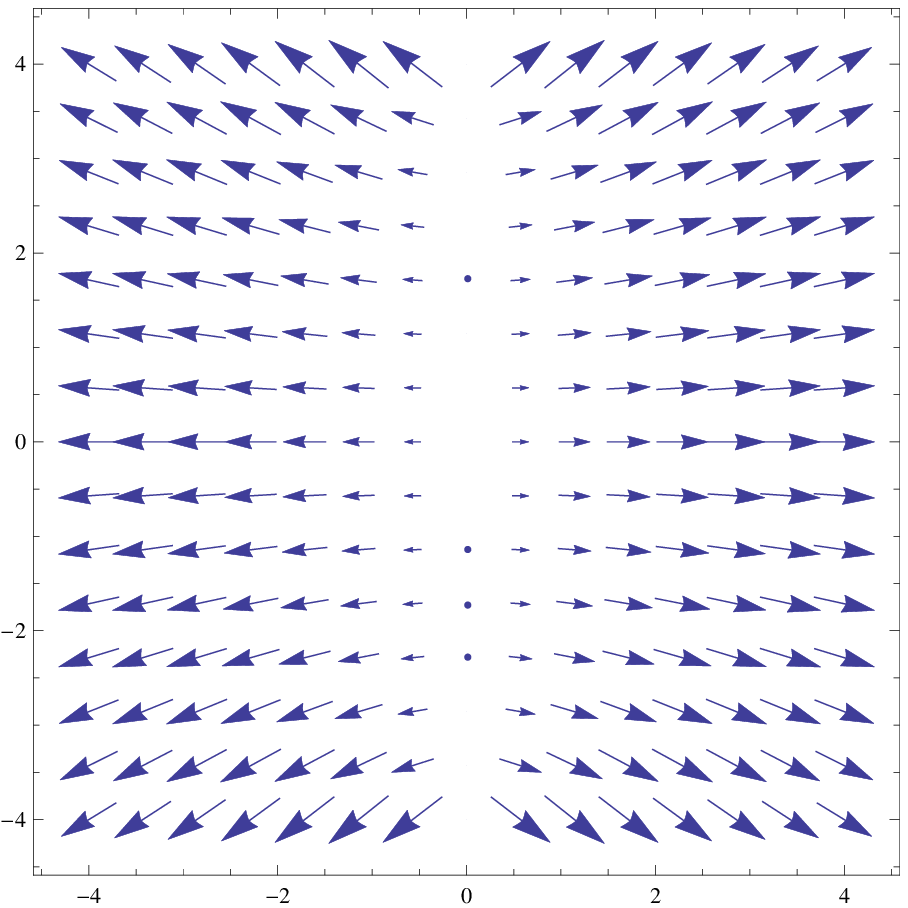}
  \includegraphics[width=70mm,height=70mm]{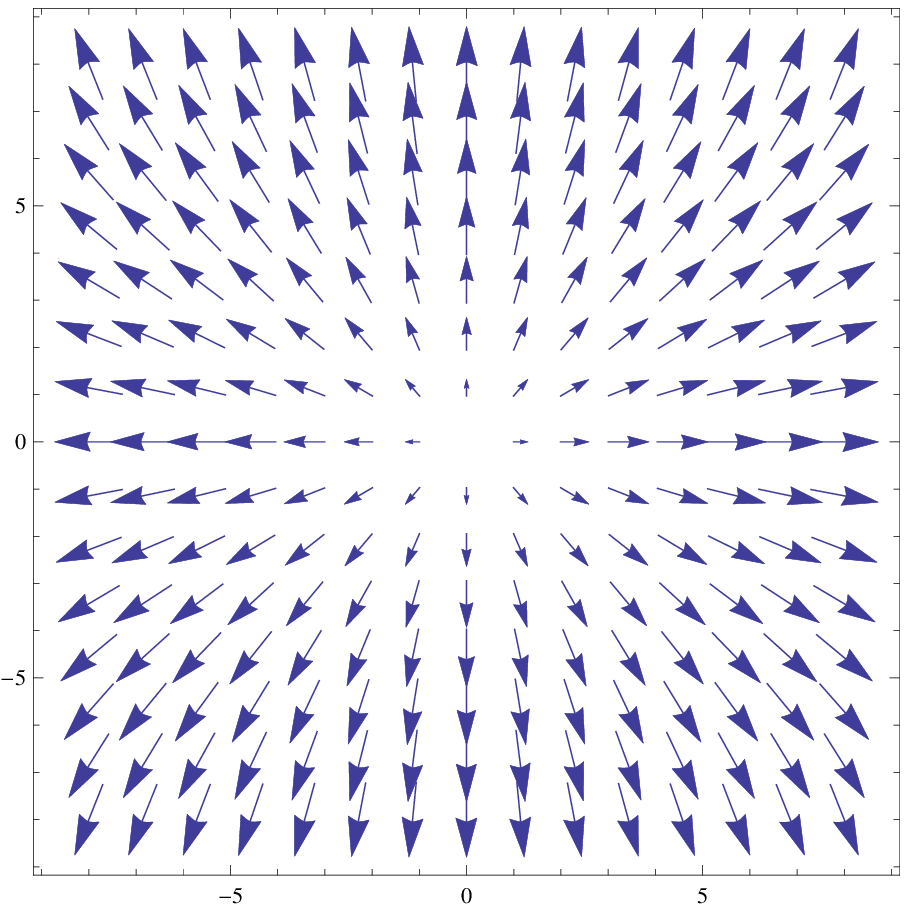}
 \caption{The flow velocity $(v_z,v_x)$ in the $(z,x)$ plane for $L=4$. Left: the early time solution (\ref{real}) with $t=1$. (As shown in Section~\ref{xt}, (\ref{real}) is valid also for $x_\perp>t$.)  Right: the late time solution (\ref{late}) with $t=10$.  \label{fig2}}
\end{figure}

For this purpose, we first observe that there  are two simple  solutions of (\ref{pot})
\beq
\hat{T}=e^{\rho_\perp}\,, \qquad \alpha=-\eta_\perp\,, \qquad \Phi=-e^{\rho_\perp}\cosh\eta_\perp\,, \label{former}\\
\hat{T}=e^{-\rho_\perp}\,, \qquad \alpha=\eta_\perp\,, \qquad \Phi=e^{-\rho_\perp}\cosh\eta_\perp\,. \label{latter}
\eeq
Substituting these into (\ref{hydro1}),
%or more explicitly,
%\beq
%\partial_{\rho_\perp} (\hat{T}^3 \cosh\rho_\perp \sinh\eta_\perp \cosh\alpha)+\partial_{\eta_\perp} (\hat{T}^3\cosh\rho_\perp \sinh\eta_\perp \sinh\alpha)=0\,, \label{sec}
%\eeq
we find that
 (\ref{former}) and (\ref{latter}) approximately solve (\ref{hydro1}) up to terms of order ${\mathcal O}(e^{\pm 2\rho_\perp})$ in the infinite 'past' $\rho_\perp \to -\infty$ and infinite `future' $\rho_\perp\to \infty$, respectively. It is tempting to regard these solutions as the asymptotic behaviors of a single solution in the limits $\rho_\perp\to \mp \infty$. Such a solution would be an attractive model of heavy-ion collisions. Indeed, in Minkowski space, the limit $\rho_\perp\to -\infty$ corresponds to early times $\tau_\perp \ll L$ and (\ref{former})  becomes
\beq
\varepsilon \propto\frac{\hat{T}^4}{\tau^4_\perp}&=&\left(\frac{2L}{\sqrt{(L^2+z^2+x_\perp^2-t^2)^2+4L^2(t^2-x_\perp^2)}
+L^2+z^2+x_\perp^2-t^2}\right)^4
\nonumber \\
&=&\left(\frac{2L}{\sqrt{(L^2+x_\perp^2-\tau^2)^2+4L^2(\tau^2\cosh^2\eta-x_\perp^2)}
+L^2+x_\perp^2-\tau^2}\right)^4\,, \label{t}
\eeq
 where in the second line we switched to the more familiar variables $\tau=\sqrt{t^2-z^2}$ and $\eta=\frac{1}{2}\ln \frac{t+z}{t-z}$ often used in heavy-ion phenomenology. We see that the solution is not boost-invariant ($\eta$-dependent) and decays exponentially at large $|\eta|$ for fixed $\tau$. Moreover, $\varepsilon=\varepsilon(\tau,\eta,x_\perp)$ depends on three variables in contrast to the Bjorken flow $\varepsilon=\varepsilon(\tau)$ and the Gubser flow $\varepsilon=\varepsilon(\tau,x_\perp)$. Note also that $\varepsilon$ is finite in the limit $\tau\to 0$.
% (In contrast, the Bjorken and Gubser flows diverge as $\varepsilon\propto 1/\tau^{4/3}$.)
  As for the flow velocity, we find\footnote{In order to obtain this result we solved the defining equation (\ref{def})
exactly for $e^{\rho_\perp}$.
However, since (\ref{former}) has been derived by neglecting terms of order $e^{2\rho_\perp}$,  one could approximate $\sinh\rho_\perp \approx -e^{- \rho_\perp}/2$. If one does this, one finds a spherical flow at early times
 \beq
 \vec{u}\approx \frac{2t\vec{r}}{L^2+r^2-t^2}\,, \nonumber
 \eeq
  where $\vec{r}=(\vec{x}_\perp, z)$.
 In order to resolve this ambiguity, one has to include the ${\mathcal O}(e^{2\rho_\perp})$ corrections. This will be discussed below.
 }

\beq
u_t&=&-\frac{1}{t^2-x_\perp^2}\left(\frac{t^2(t^2-x_\perp^2+L^2+z^2)}
{\sqrt{(t^2-x_\perp^2-L^2-z^2)^2+4L^2(t^2-x_\perp^2)}}-x_\perp^2\right),
 \nonumber \\
 \vec{u}_\perp &=& \frac{t\vec{x}_\perp}{t^2-x_\perp^2}\left(\frac{t^2-x_\perp^2+L^2+z^2}
 {\sqrt{(t^2-x_\perp^2-L^2-z^2)^2+4L^2(t^2-x_\perp^2)}} -1\right)\,, \nonumber \\
 % \nonumber \\
% = \frac{4z^2 t\vec{x}_\perp}{\sqrt{(t^2-x_\perp^2-L^2-z^2)^2+4L^2(t^2-x_\perp^2)}\left(
% t^2-x_\perp^2+L^2+z^2 + \sqrt{(t^2-x_\perp^2-L^2-z^2)^2+4L^2(t^2-x_\perp^2)}\right)} \nonumber \\
u_z&=& \frac{2zt}
{\sqrt{(t^2-x_\perp^2-L^2-z^2)^2+4L^2(t^2-x_\perp^2)}}\,. \label{real}
\eeq
Contrary to the appearance, there is no pole at $t=x_\perp$ in $u_t$ and $u_\perp$, and this suggests that the solution can be continued to $x_\perp>t$ (see Section~\ref{xt}).
The three-dimensional velocity $(v_z,\vec{v}_\perp)\equiv (-u_z/u_t,-\vec{u}_\perp/u_t)$ is plotted in Fig.\ref{fig2}(left).
The flow is expanding dominantly longitudinally $|v_z|\gg |v_\perp|$, which is what one would expect in the early stages of heavy-ion collisions.
%$|v_z|$ has a peak at
%\beq
%z=\pm \sqrt{L^2+t^2-x_\perp^2}\,.
%\eeq

The other limit $\rho_\perp \to \infty$ corresponds in Minkowski space to the late time regime $\tau_\perp \gg L$ and (\ref{latter}) becomes
\beq
\varepsilon\propto\left(\frac{2L}{t^2-x_\perp^2-z^2-L^2+
\sqrt{(t^2-x_\perp^2-z^2-L^2)^2+4L^2(t^2-x_\perp^2)}}\right)^4\,,
\label{lan}
\eeq
\beq
u_t&=&-\frac{1}{t^2-x_\perp^2}\left(\frac{t^2(t^2-x_\perp^2+L^2+z^2)}
{\sqrt{(t^2-x_\perp^2-L^2-z^2)^2+4L^2(t^2-x_\perp^2)}}+x_\perp^2\right)\,,
 \nonumber \\
 \vec{u}_\perp &=& \frac{t\vec{x}_\perp}{t^2-x_\perp^2}\left(\frac{t^2-x_\perp^2+L^2+z^2}
 {\sqrt{(t^2-x_\perp^2-L^2-z^2)^2+4L^2(t^2-x_\perp^2)}} +1\right)\,,  \nonumber \\
u_z&=& \frac{2zt}
{\sqrt{(t^2-x_\perp^2-L^2-z^2)^2+4L^2(t^2-x_\perp^2)}}\,. \label{late}
\eeq
As shown in Fig.~\ref{fig2}(right), the flow is almost spherical.
In fact, this asymptotic late time regime is not reached in actual heavy-ion collisions because the system freezes out earlier, presumably when $t\sim {\mathcal O}(L)$. Yet, the transition from one-dimensional to three-dimensional expansions is expected on general grounds \cite{landau}, and it is reassuring to see this analytically.

It is  remarkable that, although the two solutions (\ref{former}) and (\ref{latter}) are  trivially related by the reflection symmetry $\rho_\perp \to -\rho_\perp$ in $dS_2\times H_2$ (see Fig.~\ref{fig1}), they appear quite distinct in Minkowski space.
% and qualitatively mimic the spacetime evolution of heavy-ion collisions at early and late times, respectively.
In fact, similar comments apply to Gubser's solution whose  flow velocity reads, in our notation,
\beq
(v_z,v_\perp)=\left(\frac{z}{t},\frac{x_\perp}{t}\frac{2\tau^2}{L^2+\tau^2+x_\perp^2}\right)\,.
\eeq
The flow is confined within the light cone $|z|\le t$ and there is a singularity at $|z|=t$ where $T$ diverges and $|v_z|$ approaches unity irrespective of the value of $x_\perp$. Our solution has support also at $|z|>t$ and has no singularity on the light-cone $t=|z|$.\footnote{The support property of $\varepsilon$ and its physical interpretation is subject to the choice of $t_0$ in (\ref{note}). But in any case, clearly the light-cone $t=|z|$ plays no role in our solution.} It is thus more relevant to low energy collisions. Yet, our solution is qualitatively different also from the Khalatnikov-Landau 1+1-dimensional solution. The rapidity dependence in the one-dimensional stage (\ref{t}) is exponential at large $|\eta|$
 \beq
\sigma \sim e^{-3|\eta|}\,, \qquad \varepsilon \sim e^{-4|\eta|}\,, \label{expo}
 \eeq
 rather than the Gaussian-like distribution $\sigma\sim e^{\sqrt{\#-\eta^2}}$ \cite{landau,Wong:2008ex}. Moreover, in the three-dimensional stage at late times the energy density decreases as (see (\ref{lan}))
\beq
\varepsilon \sim \frac{1}{t^8}\,,
\eeq
in contrast to the $\varepsilon \sim 1/t^4$ behavior estimated by Landau. By dealing with the fully 1+3-dimensional problem already in the early stage, we have arrived at a genuinely new type of solution.

\subsection{Perturbative expansion}

Unfortunately, we have not found an exact analytical expression  $\alpha(\rho_\perp,\eta_\perp)$ which interpolates the limiting solutions (\ref{former}), (\ref{latter}). We can however construct an approximate solution perturbatively  in the form
\beq
\hat{\varepsilon}=e^{\pm 4\rho_\perp}\left(1+\sum_{k=1}^\infty a_k(\eta_\perp) e^{\pm 2k\rho_\perp}\right)\,, \quad
\alpha=\pm \left(-\eta_\perp + \sum_{k=1}^\infty b_k(\eta_\perp)e^{\pm 2k \rho_\perp}\right)\,, \quad (\rho_\perp \to \mp \infty)\,, \label{per}
\eeq
where we opt to solve in terms of $\hat{\varepsilon}$ instead of $\hat{T}$. The coefficients $a_k$, $b_k$  can be determined order by order by substituting (\ref{per}) into (\ref{hydro1}) and (\ref{pot}), or more explicitly,
\beq
\!\!\!\!\!\!3\left(\cosh\alpha\partial_{\rho_\perp} \hat{\varepsilon}  + \sinh\alpha\partial_{\eta_\perp}\hat{\varepsilon} \right) +4\hat{\varepsilon}
\left( \tanh\rho_\perp \cosh\alpha \! +\!\coth\eta_\perp \sinh\alpha\!+\! \sinh\alpha \partial_{\rho_\perp} \alpha  + \cosh\alpha \partial_{\eta_\perp}\alpha\right)\!=\!0\,,
\label{hyd2}
\eeq
\beq
\sinh\alpha \partial_{\rho_\perp}\hat{\varepsilon} +\cosh\alpha \partial_{\eta_\perp} \hat{\varepsilon} + 4\hat{\varepsilon}(\cosh\alpha \partial_{\rho_\perp}\alpha + \sinh\alpha \partial_{\eta_\perp}\alpha)=0\,. \label{hyd}
\eeq

Let us make several general remarks about the structure of this perturbative expansion.
(i) We take $a_k$ and $b_k$ to be common functions in the two regimes $\rho_\perp>0$ and $\rho_\perp<0$ (i.e., $a_k^+=a_k^-$). The solution is then invariant under the global `time' reversal $\rho_\perp \to -\rho_\perp$, $\hat{\varepsilon}\to \hat{\varepsilon}$, $\alpha\to -\alpha$, which is a property of the ideal hydrodynamic equation.\footnote{This in particular implies that $\alpha$ has to vanish when $\rho_\perp=0$, which is difficult to see from (\ref{per}) (though of course (\ref{per}) breaks down when $\rho_\perp\to 0$). Instead, we may unify the two regimes $\rho_\perp \gtrless 0$ and expand more symmetrically as
\beq
\hat{\varepsilon}=e^{\pm 4\rho_\perp}\left(1+\sum_{k=1}^\infty \frac{\tilde{a}_k(\eta_\perp)}{\cosh^{2k}\rho_\perp}\right)\,, \qquad \alpha = \eta_\perp \tanh \rho_\perp \left(1-\frac{1}{\eta_\perp}\sum_{k=1}^\infty \frac{\tilde{b}_k(\eta_\perp)}{ \cosh^{2k}\rho_\perp}\right)\,.
\eeq
This is a very complicated reorganization of the series (\ref{per}) and is not a consistent expansion if one truncates the sum to any fixed order. However, it makes  the property $\alpha(\rho_\perp=0)=0$ manifest. One can match the coefficients $\tilde{a}_k$ and $\tilde{b}_k$ with the ones in (\ref{per}). For instance, at $k=1$ we have
\beq
\tilde{a}_1= \frac{d_1-1}{6} -\frac{1}{6}\left(5d_1+1\right) \cosh2\eta_\perp  \,, \qquad \tilde{b}_1=-\frac{1}{2}\eta_\perp + \frac{d_1}{4} \sinh2\eta_\perp\,,
\eeq
where $d_1$ is the same as in (\ref{free}).
}
(ii)  Eqs.~(\ref{hyd2}) and (\ref{hyd}) are invariant under the sign flip $\eta_\perp\to -\eta_\perp$ provided $\hat{\varepsilon}$ and $\alpha$ are even and odd functions of $\eta_\perp$, respectively. This is indeed what comes out of the calculation.  Geometrically, the sign flip $\eta_\perp\to -\eta_\perp$ corresponding to jumping onto the other branch of the hyperbolic space $H_2$ (see Fig.~\ref{fig1}), though in practice $\eta_\perp$ is positive by definition. (iii) From the analysis of the first few orders of the expansion (\ref{per}), we noticed that $a_k$ and $b_k$ can be written as a linear combination of $\cosh 2k' \eta_\perp$ and $\sinh 2k' \eta_\perp$ with $k'\le k$, respectively. This implies that in practice the expansion parameter is not $e^{\pm 2\rho_\perp}$ but rather
\beq
e^{\pm 2\rho_\perp} \cosh^2\eta_\perp
%\approx \frac{L^2t^2}{(t^2-L^2-r^2)^2}
\approx \begin{cases} t^2/L^2  \quad (t\ll L)\,,\\
L^2/t^2 \quad (t\gg L)\,.
\end{cases}
\eeq
Thus the expansion breaks down when $t\sim L$, and in this intermediate regime the solution can be constructed only numerically. Note that, if we shift the initial time as in (\ref{note}), the expansion parameter is $(t-t_0)^2/L^2$.
(iv) At each order of perturbation theory, one free parameter appears as an integration constant. Thus $a_k$ and $b_k$ contain $k$ free parameters which in principle can be determined from the initial condition.\footnote{The initial condition cannot be literally set at $\rho_\perp = -\infty$ because $\hat{T}$ vanishes there. The initial $\rho_\perp$ has to be  large but finite, and there are in principle infinitely many free parameters to specify the initial condition. }

Now let us show the $k=1$ solution which is relatively simple
%\beq
%a_1=-\frac{2}{5}(2+C) + 2C \cosh2\eta_\perp\,, \qquad b_1=-\frac{1}{5}(1+3C)\sinh2\eta_\perp\,, \label{free}
%\eeq
\beq
a_1=\frac{2}{3}(d_1-1) -\frac{2}{3}(5d_1+1)\cosh2\eta_\perp\,, \qquad b_1=d_1\sinh2\eta_\perp\,, \label{free}
\eeq
where $d_1$ is a free parameter. A particularly interesting choice is $d_1=0$ in which case the correction to $\alpha$ vanishes and the flow velocity (\ref{real}) (and therefore Fig.~\ref{fig2}(left)) is unmodified to this order.  As $d_1$ is increased from zero, the flow becomes rounder, and for negative $d_1$ the transverse flow changes the directions. For phenomenological purposes one can simply set $d_1$ to be zero.
%Another interesting choice is $d_1=1$ which we shall consider in the next section
%\beq
%a_1=-4\cosh2\eta_\perp\,, \qquad b_1=\sinh 2\eta_\perp\,. \label{an}
%\eeq
The $k=2$ solution is more involved
\beq
a_2&=&\frac{2}{107}\left[37+33d_2-6d_1+84d_1^2 +(38-76d_2+150d_1+40d_1^2)\cosh2\eta_\perp +107d_2\cosh4\eta_\perp\right]\,,\\
b_2&=&\frac{\sinh2\eta_\perp}{321}\left[17+180d_2-(130+427d_1)d_1+
\bigl(20-468d_2+(338+775d_1)d_1\bigr)\cosh2\eta_\perp
\right]\,,
\eeq
 where $d_2$ is the new free parameter which appears at this order.
 In the spirit that the correction to $\alpha$ is made as small as possible, we may choose $d_1=0$ and $d_2=\frac{5}{117}$ in which case
  \beq
 a_2=\frac{2}{117}\left(42+38\cosh 2\eta_\perp +5\cosh 4\eta_\perp\right)\,, \qquad  b_2=\frac{\sinh2\eta_\perp}{13}\,. \label{choice}
 \eeq
% There are specific  values of $d_1,d_2$ which make the coefficients drastically simple. For instance, setting $d_1=0$ and $d_2=1/2$ we find
%  \beq
% a_2= 1+\cosh4\eta_\perp\,, \qquad b_2=\frac{\sinh2\eta_\perp -\sinh4\eta_\perp}{3}\,.
% \eeq

Starting from $k=3$, the equations become quite complicated and the coefficients of $\cosh k'\eta_\perp$ and $\sinh k'\eta_\perp$ ($k'\le k$) tend to become very large numbers. Rather than writing down the most general result which is not illuminating, here we only show the special solution obtained for $d_1=0$, $d_2=\frac{5}{117}$, and a particular value of the new parameter at $k=3$ which makes the coefficient of $\sinh6\eta_\perp$ in $b_3$ vanish
\beq
a_3&=& -\frac{2}{118989}\left(43573+42504\cosh2\eta_\perp + 5910\cosh4\eta_\perp + 355\cosh6\eta_\perp \right)\,,\nonumber \\
b_3&=& \frac{1}{4407} \left(-282\sinh2\eta_\perp +43\sinh4\eta_\perp\right)\,.
\eeq
%\beq
%a_3&=&  -\frac{1}{1215}\left(950+840 \cosh 2\eta_\perp + 444 \cosh 4 \eta_\perp +554 \cosh 6\eta_\perp \right)\,, \\
%b_3&=&  -\frac{1}{270}\left(25 \sinh 2 \eta_\perp +2\sinh 4 \eta_\perp - 7\sinh 6 \eta_\perp \right)\,.
%\eeq
In principle, we can continue this procedure to arbitrary higher orders. We however stop here because in the next section we shall see that already the $k=2$ terms are subleading compared with the viscous correction.

%On the other hand,
%the restriction on $x_\perp$ is actually milder because, as we demonstrated above, we can eliminate the largest term in $\alpha$ at each order
%\beq
%\sinh 2k \eta_\perp e^{\pm 2k \rho_\perp} \sim \sinh \eta_\perp \cosh^{2k-1}\eta_\perp e^{\pm 2k\rho_\perp}\sim \frac{t^{2k-1}x_\perp}{L^{2k}}\,.
%\eeq
%This actually allows

\subsection{Solution at $x_\perp>t$}
\label{xt}

By introducing the coordinates $\tau_\perp,\eta_\perp$, so far we have implicitly assumed that $t>x_\perp$. However, as we noted already, the flow profile (\ref{real}) at early times has no singularity at $t=x_\perp$, and this suggests that the solution can be smoothly continued to $x_\perp>t$. To show that this is indeed the case, define for $x_\perp>t$
\beq
\tilde{\tau} = \sqrt{x_\perp^2-t^2}\,,\qquad \tilde{\eta} = \frac{1}{2}\ln \frac{x_\perp+t}{x_\perp-t}\,.
\eeq
We can then write
\beq
ds^2=-dt^2+dx_\perp^2 + x_\perp^2 d\phi^2+dz^2 &=& d\tilde{\tau}^2 -\tilde{\tau}^2d\tilde{\eta}^2 +\ttau^2\cosh^2\teta d\phi^2 + dz^2\nonumber \\
&=& \ttau^2\left(-d\teta^2+\cosh^2\teta d\phi^2+ \frac{d\ttau^2 + dz^2}{\ttau^2}\right) \nonumber \\
&=& \ttau^2\left(-d\teta^2+\cosh^2\teta d\phi^2+ d\tilde{\rho}^2 + \sinh^2 \tilde{\rho} d\tilde{\Theta}^2\right)\,,
\eeq
 where
 \beq
 \cosh\tilde{\rho}= \frac{L^2+\tilde{\tau}^2+z^2}{2L\tilde{\tau}}\,,\qquad \sinh\tilde{\rho}=\pm \frac{\sqrt{4z^2L^2 +(L^2-\tilde{\tau}^2-z^2)^2}}{2L\tilde{\tau}}\,, \qquad
 \tan \tilde{\Theta} = \frac{L^2-\tilde{\tau}^2-z^2}{2Lz}\,. \label{trans}
 \eeq
The Weyl-transformed space $ds^2/\tilde{\tau}^2$ is again $dS_2\times H_2$. The $\pm$ sign in (\ref{trans}) reflects the fact that the hyperbolic space $H_2$ consists of two disconnected spaces.
%An exact solution analogous to (\ref{ex}) is
%\beq
%\hat{u}^{\tilde{\eta}}=1, \qquad \hat{\epsilon}=\left(\frac{1}{\cosh \tilde{\eta}}\right)^{4/3}
%\eeq
Noting that $\tilde{\eta}$ is now the time-like variable, we look for solutions of the form
\beq
(\hat{u}^{\tilde{\eta}},\hat{u}^\phi, \hat{u}^{\tilde{\rho}},\hat{u}^\theta)=(\cosh\tilde{\alpha},0,\sinh\tilde{\alpha},0)\,.
\eeq
The (approximate) solution which matches with the previous solution (\ref{former}) at $t=x_\perp$ is is obtained in the case $\tilde{\rho}>0$ and reads
\beq
\hat{T}=e^{-\tilde{\rho}}\,, \qquad \tilde{\alpha}=\tilde{\eta}\,, \qquad (\tilde{\rho}\to \infty)\,.
%\hat{T}=e^{\tilde{\rho}}\,, \qquad \tilde{\alpha}=-\tilde{\eta}\,, \qquad (\tilde{\rho}\to -\infty)\,.
\eeq
One can again develop a perturbation theory around this solution. The solution in Minkowski space is identical to (\ref{t}) and (\ref{real}), and now we see that these results are valid both for $t>x_\perp$ and $t<x_\perp$.

\section{Viscous corrections}

Finally in this section, we study the effect of viscosity. The shear viscosity $\xi$ enters the nonequilibrium part of the energy momentum tensor
\beq
\delta \hat{T}^{\mu\nu}=-2\xi \hat{\sigma}^{\mu\nu}\,,
\eeq
 where $\hat{\sigma}^{\mu\nu}$ is the shear tensor.  In a conformal theory, $\xi$ scales as
\beq
\xi\propto \hat{\varepsilon}^{3/4} \sim e^{\pm 3\rho_\perp}\,.
\eeq
The hydrodynamic equation (\ref{eq}) is modified as
\beq
\hat{u}^\mu \partial_\mu \hat{\varepsilon}+\frac{4\hat{\varepsilon}}{3}\nabla_\mu \hat{u}^\mu &=&2\xi \hat{\sigma}^{\mu\nu}\hat{\sigma}_{\mu\nu}\,, \label{last} \\
4\hat{\varepsilon}\hat{u}^\nu \nabla_\nu \hat{u}^\mu+\Delta^{\mu\nu}\partial_\nu \hat{\varepsilon} &=& 6\Delta^\mu_{\,\nu}\nabla_\lambda (\xi\hat{\sigma}^{\lambda\nu})\,.
\label{mm}
\eeq
Let us compute $\hat{\sigma}^{\mu\nu}$ for the $k=1$ solution (\ref{free})
 \beq
 \alpha=\pm \left(-\eta_\perp + d_1 \sinh 2\eta_\perp e^{\pm 2\rho_\perp}\right)\,.\label{term}
 \eeq
 The nonvanishing components are found to be
\beq
\hat{\sigma}^{\rho_\perp \rho_\perp}&\approx&
%\mp \frac{1}{3}\cosh\eta_\perp \sinh^2 \eta_\perp (1\pm \tanh\rho_\perp) \approx
\mp \frac{2}{3}(1-d_1)\cosh\eta_\perp \sinh^2 \eta_\perp e^{\pm 2\rho_\perp}\,,  \qquad
\hat{\sigma}^{\rho_\perp \eta_\perp}\approx\frac{2}{3}(1-d_1)\cosh^2\eta_\perp \sinh \eta_\perp e^{\pm 2\rho_\perp} \,, \nonumber \\
\hat{\sigma}^{\Theta\Theta}&\approx&  \pm \frac{4}{3}(1-d_1)\frac{\cosh\eta_\perp}{\cosh^2 \rho_\perp}e^{\pm 2\rho_\perp}\,, \qquad
\hat{\sigma}^{\phi\phi}\approx\mp \frac{2}{3}(1-d_1)\frac{\cosh\eta_\perp}{\sinh^2\eta_\perp}e^{\pm 2\rho_\perp}\,,\nonumber \\
\hat{\sigma}^{\eta_\perp \eta_\perp}&\approx& \mp \frac{2}{3}(1-d_1)\cosh^3\eta_\perp e^{\pm 2\rho_\perp}\,. \label{sigma}
\eeq
Due to a cancelation,  the leading term is $\hat{\sigma}^{\mu\nu}\sim {\mathcal O}(e^{\pm 2\rho_\perp})$, and the two terms in (\ref{term}) are equally important. In particular, the special value $d_1=1$ makes $\hat{\sigma}^{\mu\nu}$ vanish to this order. In fact, irrespective of the value of $d_1$, it is necessary to also retain the $k=2$ terms in $\alpha$ which give ${\mathcal O}(e^{\pm 4\rho_\perp})$ contributions to $\hat{\sigma}^{\mu\nu}$. To explain this,  note that the right hand side of (\ref{last}) is ${\mathcal O}(e^{\pm 7\rho_\perp})$, while that of (\ref{mm}) is naively  ${\mathcal O}(e^{\pm 5\rho_\perp})$, but actually the coefficient of $e^{\pm 5\rho_\perp}$ vanishes. The leading contribution is then ${\mathcal O}(e^{\pm 7\rho_\perp})$, and this comes from the ${\mathcal O}(e^{\pm 4\rho_\perp})$ corrections to $\hat{\sigma}^{\mu\nu}$ as well as the ${\mathcal O}(e^{\pm 2\rho_\perp})$ corrections to the shear viscosity
%\beq
%\xi= \xi_0\hat{\varepsilon}^{3/4}\approx \xi_0 e^{\pm 3\rho_\perp} \left(1-\cosh^2\eta_\perp e^{\pm 2\rho_\perp}\right)\,,
%\eeq
\beq
\xi= \xi_0\hat{\varepsilon}^{3/4}\approx \xi_0 e^{\pm 3\rho_\perp} \left(1+\frac{1}{2}\Bigl(d_1-1 -(5d_1+1)\cosh 2\eta_\perp \Bigr)e^{\pm 2\rho_\perp}\right) \,,
\eeq
where $\xi_0$ is a constant.
  To the order of interest, (\ref{last}) and (\ref{mm}) reduce to
\beq
\!\!3\left(\cosh\alpha\partial_{\rho_\perp} \hat{\varepsilon}  + \sinh\alpha\partial_{\eta_\perp}\hat{\varepsilon} \right) &+&4\hat{\varepsilon}
\left( \tanh\rho_\perp \cosh\alpha + \coth\eta_\perp \sinh\alpha+ \sinh\alpha \partial_{\rho_\perp} \alpha  + \cosh\alpha \partial_{\eta_\perp}\alpha\right) \nonumber \\
&=&16(1-d_1)^2\xi_0 \cosh^2\eta_\perp e^{\pm 7\rho_\perp}\,,
\eeq
%\beq
%\sinh\alpha \partial_{\rho_\perp}\hat{\varepsilon} +\cosh\alpha \partial_{\eta_\perp} \hat{\varepsilon} &+& 4\hat{\varepsilon}(\cosh\alpha \partial_{\rho_\perp}\alpha + \sinh\alpha \partial_{\eta_\perp}\alpha)=6\cosh\alpha \nabla_\lambda(\xi \hat{\sigma}^{\lambda\eta_\perp})-6\sinh\alpha \nabla_\lambda (\xi\hat{\sigma}^{\lambda\rho_\perp}) \nonumber \\ &=&  \pm 16 \xi_0 \sinh\eta_\perp \cosh\eta_\perp e^{\pm 7\rho_\perp}\,.
%\eeq
\beq
\sinh\alpha \partial_{\rho_\perp}\hat{\varepsilon} +\cosh\alpha \partial_{\eta_\perp} \hat{\varepsilon} + 4\hat{\varepsilon}(\cosh\alpha \partial_{\rho_\perp}\alpha + \sinh\alpha \partial_{\eta_\perp}\alpha)
%&=&6\cosh\alpha \nabla_\lambda(\xi \hat{\sigma}^{\lambda\eta_\perp})-6\sinh\alpha \nabla_\lambda (\xi\hat{\sigma}^{\lambda\rho_\perp}) \nonumber \\
=  \pm C\xi_0 \sinh 2 \eta_\perp e^{\pm 7\rho_\perp}\,,
\eeq
where
\beq
C\equiv \frac{24}{107}(25-576d_2+88d_1+319d_1^2)\,.
\eeq
The solution
to linear order in $\xi_0$ is
\beq
\hat{\varepsilon}&\approx& e^{\pm 4\rho_\perp} +\frac{2}{3} \Bigl(d_1-1-(5d_1+1)\cosh2\eta_\perp\Bigr) e^{\pm 6\rho_\perp}\pm \xi_0 A(\eta_\perp)e^{\pm 7\rho_\perp}\,, \nonumber \\
 \alpha&\approx&\pm \Bigl(-\eta_\perp + d_1 \sinh 2\eta_\perp e^{\pm 2\rho_\perp}\pm  \xi_0 B(\eta_\perp)e^{\pm 3\rho_\perp} \Bigr)\,, \label{solution}
\eeq
 where
\beq
A(\eta_\perp)&=& \frac{1}{3}\left(\frac{4(1-d_1)^2}{3} -\frac{88D}{21}\pm \frac{5C}{14}\right)\cosh 3\eta_\perp +\left(\frac{4(1-d_1)^2}{3}+\frac{8D}{21}\mp\frac{3C}{14}\right)
\cosh\eta_\perp\,, \nonumber \\
B(\eta_\perp)&=& D \sinh\eta_\perp \cosh2\eta_\perp + \frac{1}{7}\left(D\pm \frac{3C}{4}\right)\sinh\eta_\perp\,,
\eeq
 and $D$ is an arbitrary constant. We see that  the viscous effect brings in odd powers of $e^{\pm \rho_\perp}$ in the series and is parametrically  larger than the $k=2$ corrections. Nevertheless, $A,B$ are sensitive to the $k=2$ flow velocity via the constant $C$.
This is not inconsistent because we treat $\xi_0$ as a small parameter and keep only the linear terms in $\xi_0$. The viscous effect may modify the $k=2$ solution, but the backreaction of this onto the solution (\ref{solution}) via $\hat{\sigma}^{\mu\nu}$ is of higher order in $\xi_0$ and can be neglected.
 Note that the $A,B$-terms in (\ref{solution}) manifestly break the symmetry under $\rho_\perp\to - \rho_\perp$. This is because viscosity breaks the time-reversal symmetry of the hydrodynamic equations. As a consequence, the entropy is not conserved 
 \beq
 \nabla_\mu (\sigma u^\mu)= \frac{16\xi_0}{3}(1-d_1)^2 \cosh^2\eta_\perp e^{\pm 6\rho_\perp} + {\mathcal O}(e^{\pm 8\rho_\perp})\,.
 \eeq

\begin{figure}[tbp]
  \includegraphics[width=52mm,height=52mm]{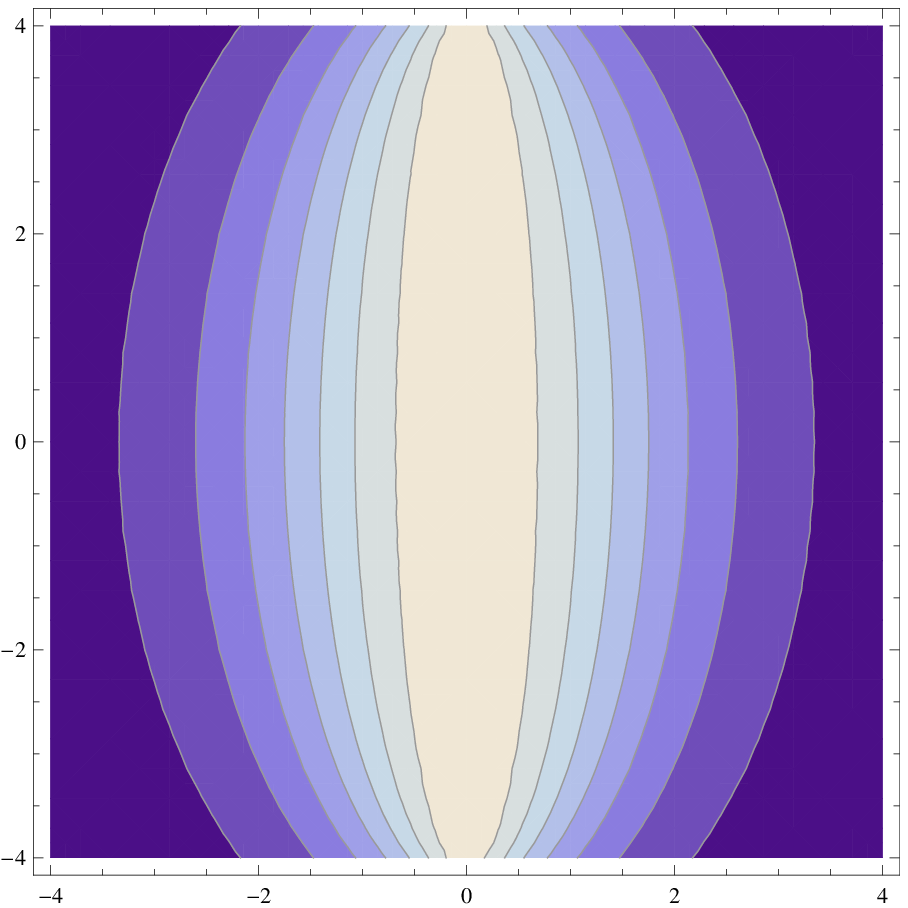}
  \includegraphics[width=52mm,height=52mm]{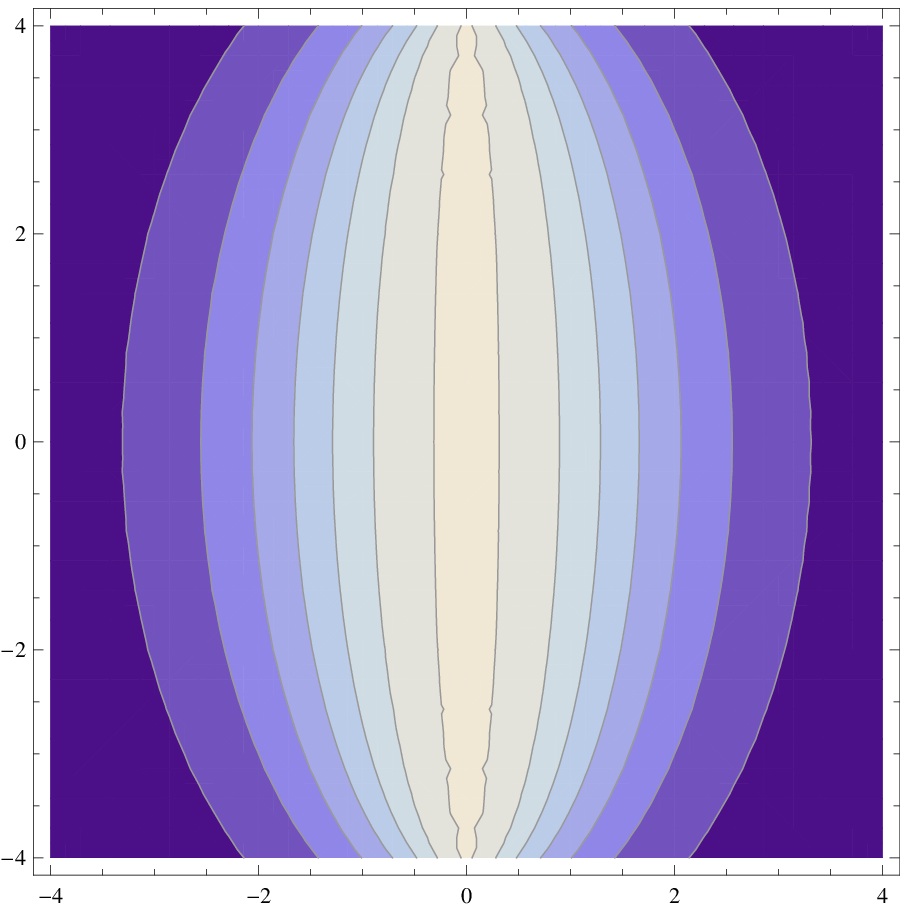}
   \includegraphics[width=52mm,height=52mm]{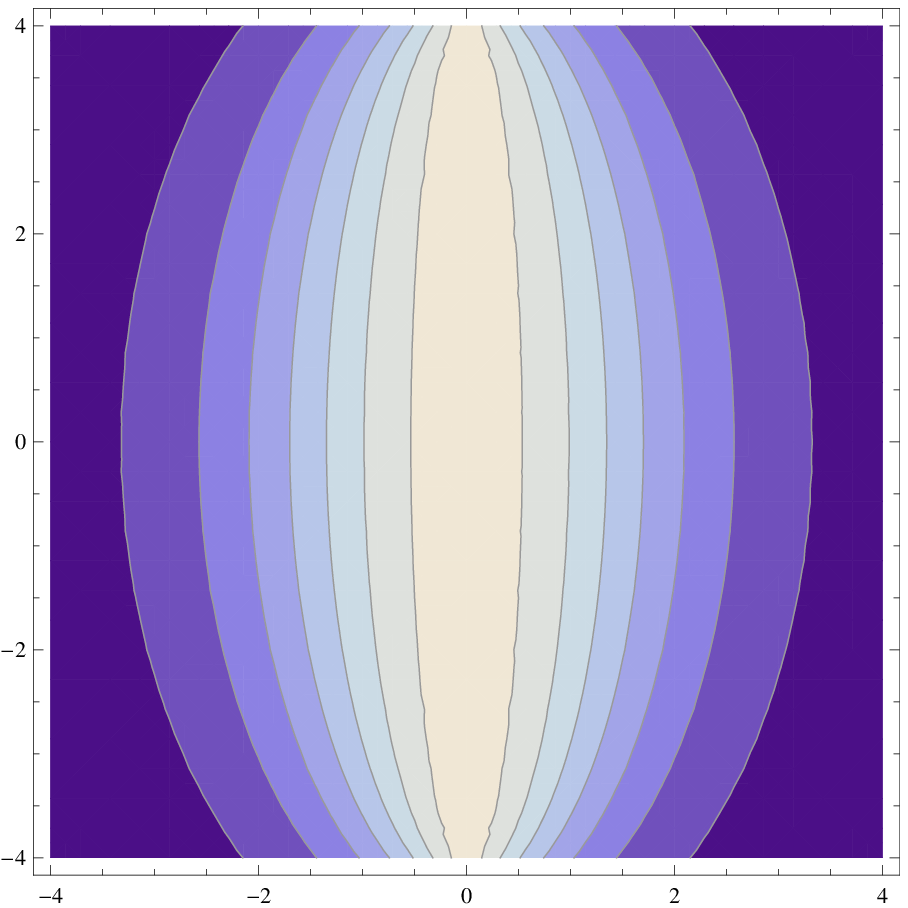}
 \caption{The contour plot of the energy density $\varepsilon$ in the $(z,x)$ plane. Left: the first term of (\ref{solution}). Middle: the first and second terms of (\ref{solution}). Right: all terms of (\ref{solution}) included. We have set $t=1$ and $L=4$ as in Fig.~\ref{fig2}. For an illustrative purpose, we used a somewhat large value $\xi_0=1.5$.  \label{fig3}}
\end{figure}

To see the effect of viscosity quantitatively, as an illustration let us choose $d_1=0$ and $d_2=\frac{5}{117}$ as was done in (\ref{choice}). We furthermore set $D=0$ to eliminate the $\sinh3\eta_\perp$ term in $B(\eta_\perp)$. Then $C=\frac{120}{1391}$ is negligibly small and the flow is still dominantly longitudinal at early times $\alpha \approx - \eta_\perp$. The energy density becomes
\beq
\varepsilon\approx \frac{1}{\tau_\perp^4}\left(e^{4\rho_\perp} -\frac{4}{3}\cosh^2\eta_\perp e^{ 6\rho_\perp}+ \xi_0\frac{16}{9}\cosh^3\eta_\perp e^{ 7\rho_\perp}\right)
= T^4\left(1-\frac{4}{3}t^2 T^2+\frac{16\xi_0}{9}t^3T^3\right)\,, \label{solution}
\eeq
where $T$ is as in (\ref{t}).
This is plotted in Fig.~\ref{fig3}. The left and middle figures show that the first correction ($k=1$) makes the flow more anisotropic. The viscous effects counteracts this change and tends to make the flow rounder, consistently with the general expectations. When $D$ is increased from zero, the flow velocity $u^\mu$ becomes relatively more isotropic. At the same time, $A$ is reduced, and so is the viscous effect on $\varepsilon$.\\

 In conclusion, we have presented a novel solution of 1+3-dimensional relativistic hydrodynamics which essentially depends on three variables $\tau,\eta,x_\perp$ and qualitatively captures the salient features of the evolution of fireballs in heavy-ion collision. From our point of view, the initial one-dimensional expansion and the final three-dimensional expansion are geometrically related by the reflection symmetry $\rho_\perp \to -\rho_\perp$ in the associated $dS_2$ space.  Our non-boost-invariant solution is essentially different from the boost-invariant Bjorken and Gubser solutions. Moreover, our fully 1+3-dimensional treatment has led to the exponential rapidity dependence (\ref{expo}), which makes this solution also distinct from the Khalatnikov-Landau solution.

Phenomenologically, our solution is more relevant to low energy heavy-ion collisions rather than high energy. The late time regime $\rho_\perp\to \infty$ is presumably  not reached in practice because the freezeout sets in earlier. Also, the value of $t_0$ in (\ref{note}) should be adjusted in order to mimic the initial matter distribution. We hope to return to these problems elsewhere.

\section*{Acknowledgements}

D.~Y. is supported by the RIKEN Foreign Postdoctoral Researcher program.
Y.~H. thanks the Center of Nuclear Matter Science of Central China Normal University for hospitality when this work was completed.

\end{document}